\documentclass[10pt]{amsart}

\usepackage{amsmath}
\usepackage{amssymb}

\usepackage{graphicx}

\usepackage{cite}

\usepackage{color} 

\usepackage[labelfont=bf,labelsep=period,justification=raggedright]{caption}

\makeatletter
\renewcommand{\@biblabel}[1]{\quad#1.}
\makeatother

\date{}

\begin{document}

\title{Comprehensive Detection of Genes Causing a Phenotype using Phenotype Sequencing and Pathway Analysis}
\author{
Marc Harper$^{1}$,
Luisa Gronenberg$^{5}$,
James Liao$^{1,5}$,
Christopher Lee$^{1,2,3,4}$
}\footnotetext[1]{ Institute for Genomics and Proteomics, University of California, Los Angeles, CA, USA }
\footnotetext[2]{ Dept. of Chemistry \& Biochemistry,  University of California, Los Angeles, CA, USA }
\footnotetext[3]{ Dept. of Computer Science,  University of California, Los Angeles, CA, USA }
\footnotetext[4]{ Molecular Biology Institute,  University of California, Los Angeles, CA, USA }
\footnotetext[5]{ Department of Chemical and Biomolecular Engineering,   University of California, Los Angeles, CA, USA }
\footnotetext[6]{E-mail: Corresponding marcharper@ucla.edu }

\maketitle

\begin{abstract}
\label{paper:abstract}\label{paper:comprehensive-discovery-of-genes-causing-a-phenotype-using-phenotype-sequencing-and-pathway-analysis}
Discovering all the genetic causes of a phenotype is an
important goal in functional genomics.  In this paper we
combine an experimental design for multiple
independent detections of the genetic
causes of a phenotype, with a high-throughput sequencing
analysis that maximizes sensitivity for comprehensively
identifying them.  Testing this
approach on a set of 24 mutant strains generated
for a metabolic phenotype with many known genetic
causes, we show that this pathway-based phenotype
sequencing analysis greatly improves
sensitivity of detection compared with previous
methods, and reveals a wide range of pathways
that can cause this phenotype.
We demonstrate our approach on a
metabolic re-engineering phenotype,
the PEP/OAA metabolic node in E. coli, which is crucial to
a substantial number of metabolic pathways and under renewed interest
for biofuel research.  Out of 2157 mutations
in these strains, pathway-phenoseq discriminated
just five gene groups (12 genes) as statistically
significant causes of the phenotype.  Experimentally,
these five gene groups, and the next two high-scoring
pathway-phenoseq groups, either have a clear connection to the PEP
metabolite level or offer
an alternative path of producing oxaloacetate (OAA),
and thus clearly explain the phenotype.  These high-scoring
gene groups also show strong evidence of positive
selection pressure, compared with strictly neutral selection
in the rest of the genome.
\end{abstract}


\section{Introduction}
\label{paper:introduction}
Discovering what genes cause a specific phenotype poses
several experimental and analytical challenges, and there
are several approaches in the literature for causal gene identification
including direct identification of causal mutations from naturally evolving
populations growing in the prescence of isobutanol \cite{minty2011evolution} 
\cite{Atsumi2011} , using transposon insertions to detect antibiotic
targets \cite{umland2012vivo} , use of chemical mutagenesis to produce
randomly generated mutants and subsequent high-throughput sequencing to
identify key mutation \cite{sarin2010analysis}  \cite{zhu2012gene} 
\cite{Harper2011} . In particular, the method described in \cite{Harper2011} ,
called phenotype sequencing, combines the last approach with sequencing
techniques to produce more information at a substantially reduced total cost.
See \cite{erlich2009dna}  and \cite{emanuele2012snp}  for more on pooling
methods.

Many such methods, while successful, have substantial drawbacks in terms of
efficiency and comprehensivity of detection, total labor required to
create mutants and verify mutations as causal, and overall cost. Unless the
mutagenesis density is very low, there can be many mutations that must be
checked; if there is only a single mutation in each mutant, causes of complex
phenotypes requiring more than one mutation may be missed. Naturally evolved
strains typically both have fewer
mutations (10-20 typically) and a larger fraction of these
directly contribute to the phenotype \cite{Honisch2004} 
\cite{Velicer2006}  \cite{Smith2008} 
\cite{Klockgether2010}  \cite{LeePalsson2010}  \cite{Chen2010} ,
with in some cases as few as 3 mutations per strain
\cite{Herring2006}  \cite{Conrad2009}  or more than
40 \cite{Srivatsan2009} . On the other hand,
natural evolution typically involves a mixture
of many mutants competing with each other.
Even small differences in selective advantage
will tend to give a winner-take-all outcome,
in which the ``top'' mutant takes over the culture,
and other causes of the phenotype are obscured.
This can occur even over a relatively short period
of competitive culture (illustrated in Fig. 1).
Hence if mutants
are allowed to compete, detection of smaller contributors to the phenotype can
be washed out by the growth of other mutants. This means that only some of the
causes of a particular phenotype will be detected.
In particular, if there is a ``trivial'' way to get
the phenotype, this can obscure the interesting,
non-obvious causes of the phenotype.

While mutagenesis can aid the production of mutants with the
desired phenotype, it also elevates the total number of mutations
in each strain (often 50 - 100 mutations \cite{Ohnishi2008}  \cite{LeCrom2009} ,
of which perhaps only one actually causes the phenotype). Dissecting
these many candidate mutations experimentally can be laborious, so we employ
statistical methods to detect which mutations are most likely to be causal.
\begin{figure}[htbp]
\centering

\includegraphics[scale=0.4]{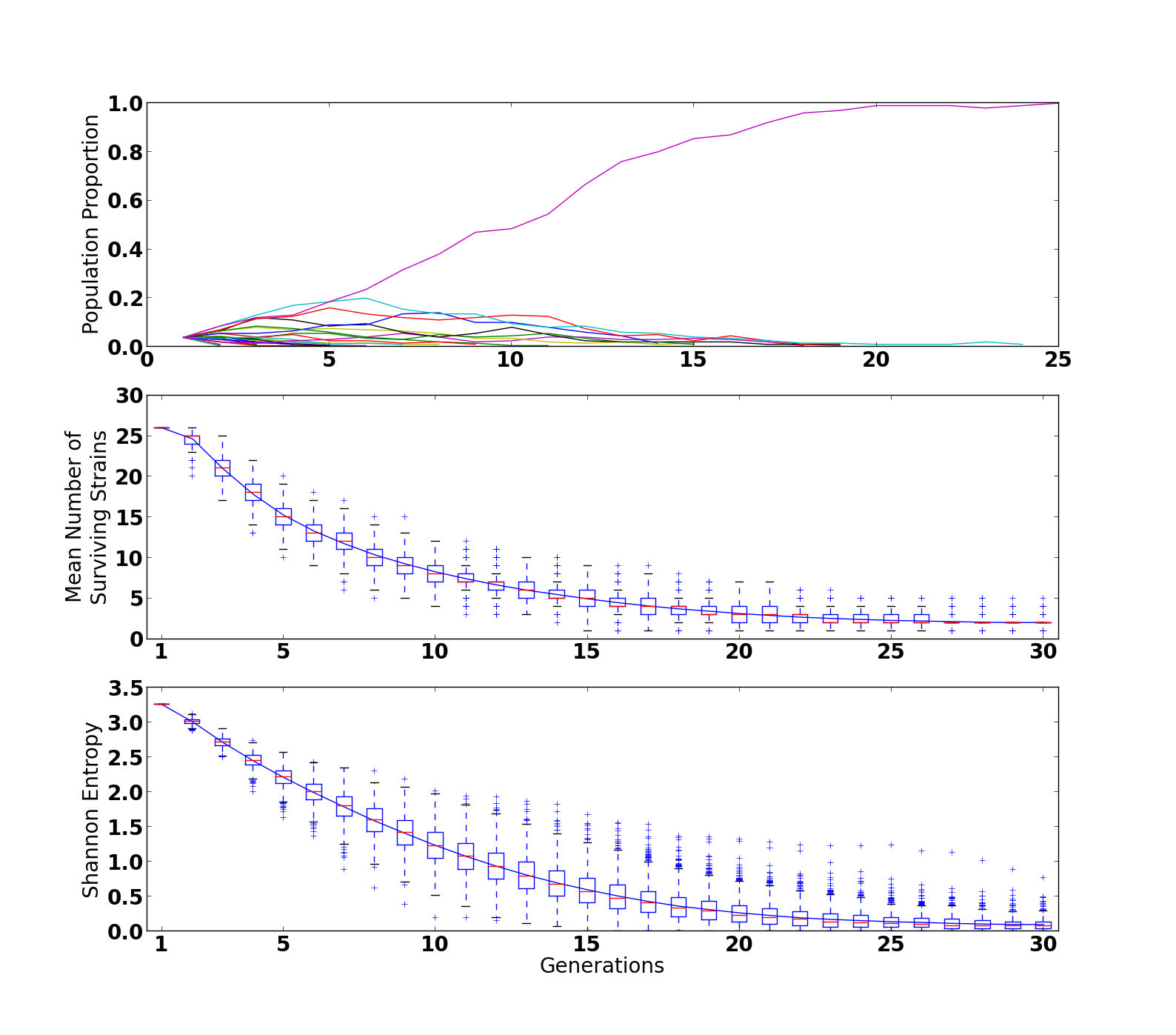}
\caption{\emph{1000 Simulations of the Wright-Fisher Selective Dynamics
\cite{ewens2004mathematical} of a Randomly Mutagenized Population.}}{\small 
\textbf{A}. \emph{(Top) a simulation of 26 strains of various fitnesses
that grow exponentially from a founder population of individual
mutants to a carrying capacity under Wright-Fisher selection dynamics.
The results of a single simulation show that one mutant
dominates the population after a small number of generations.
Note diversity is lost due not only to selection, but also
genetic drift.}

\textbf{B}. \emph{(Middle) As reproduction and selection proceeds,
the mean number of distinct strains decreases very quickly.
On average half of the strains are lost after just 6-7 generations.}

\textbf{C}. \emph{(Bottom) Similarly, the mean Shannon entropy \cite{CoverThomas91}
of the population distribution also decreases quickly. This differs
from (B) in that the population proportions are also taken
into account.}
}\end{figure}

Given these challenges, it would be very useful to
have reliable high-throughput methods for
comprehensively identifying all the genetic causes of a
phenotype.  Three features seem crucial for this goal.
First, sufficient mutagenesis coverage is required to hit
all the potential causes of the phenotype.  Note
this may require mutating two or more genes
simultaneously to achieve the desired
phenotype. For a gene to be
identified with any kind of statistical significance
in a high-throughput (genome-wide) analysis, the ``target''
set of mutations in a gene that can actually cause
the phenotype must be hit not just once but multiple
times in independent strains.
Second, the different mutant strains (representing
independent mutagenesis events) must be screened
non-competitively, e.g. by either picking only one
colony from each independent experiment, or by
forgoing long growth rescue in liquid medium to avoid
multiple colonies arising from genetically
identical daughter cells of a single mutant.
This ensures that the different strains with the phenotype
will be independent mutation events
that represent an unbiased sampling
of the diverse possible causes of the phenotype.
High-throughput sequencing of the independent
mutant strains, yielding the total number of times
a gene is independently ``hit'' by mutations across all
the strains, can then directly reveal genes that cause
the phenotype \cite{Harper2011} . We refer to this bioinformatic
approach as ``phenotype sequencing''.
The results of the first phenotype sequencing experiment were
further verified in the study by Minty et al \cite{minty2011evolution} ,
which found specific causal mutations in many of the genes identified
by phenotype sequencing (and also verified partially by \cite{Atsumi2011} ;
see also \cite{reyes2011genomic} ). We argue now that phenotype sequencing is
self-validating and also present strength of selection measures to further
statistically validate results.

Third, to attain
sensitive and \emph{comprehensive} discovery of the causal genes,
the analysis must be able to combine signals
across multiple genes that function together,
e.g. in the same pathway.  When
multiple genes in a pathway can cause the same
phenotype, this ``splits'' the signal (concretely,
the number of observed mutations) among them,
making it much harder to detect.  For example,
our first phenotype sequencing analysis
did not obtain a statistically significant score
for some genes that are known to cause the phenotype,
even though they were relatively highly ranked (due
to having more mutations than expected by random
chance) \cite{Harper2011} .  Combining signals
from multiple such genes in a pathway greatly
improves sensitivity and hence allows for comprehensive discovery.

To assess the possibility of attaining these goals,
we first developed a ``pathway-phenoseq'' analysis that
combines mutation signals across each specified pathway.
For this first test, we used pathway information
from the EcoCyc database of functionally associated genes
in \emph{E. coli} \cite{EcoCyc11} .  Second, we tested its
ability to discover the multiple genetic causes
of a metabolic phenotype which is known to involve many
genes and pathways. Third, in addition to validating
its results against the experimental literature, we
also developed bioinformatic validation methods
based on gene clustering and independent measures
of positive selection.

As an experimental test, we sought a phenotype that
involves many pathways and where existing experimental
literature could validate the results of our analysis.
We therefore chose a metabolic phenotype, namely
recovery of ability to grow on glucose by \emph{E. coli}
lacking the Phosphoenolpyruvate carboxylase (PPC).
Metabolic engineering of the pyruvate - phosphoenolpyruvate
(PEP) - oxaloacetate (OAA) node has long been studied
as a way of modifying the energy balance of the cell
\cite{ChaoLiao93}  \cite{KimLaivenieks04}  \cite{LiaoChao94} .
Phosphoenolpyruvate carboxylase (PPC) carboxylates PEP
to OAA while releasing inorganic phosphate. Phosphoenolpyruvate
carboxykinase (PEPCK) decarboxylates OAA and activates
it to PEP using ATP as a substrate. PEPCK is reversible
in some organisms and the reverse PEPCK reaction is more
energy efficient than the PPC reaction because it conserves
the phosphate from PEP by generating ATP. In \emph{E. coli},
however, PEPCK is not reversible under normal conditions.
Recently, with the increased focus on renewable resources
from microbes, optimization of the PEP/OAA metabolic
node has received renewed interest. Much of the interest
has focused on increasing the production of succinate
(a high value carboxylic acid of industrial relevance)
in \emph{E. coli} and other microbes. In recent studies PPC
has been supplemented or replaced with pyruvate carboxylase
(PYK) \cite{Blankschien10} , and more often with PEPCK
\cite{Zhang09} , which increases the ATP pool of the
cell. This can lead to higher levels of succinate production
from a variety of feedstocks \cite{Liu12} . The use of
PEPCK and the resulting higher ATP concentration has
also been exploited to increase production of malate,
OAA \cite{Park12}  or fumarate \cite{Zhang12}  and even
the amount of recombinantly expressed proteins \cite{KimKwon2012} .

Overexpressing either a native or heterologous pepck gene
is one way to compensate for the knockout of \emph{ppc} and
rescue an OAA auxotroph strain. However, a number of
other pathways affect OAA levels and flux through the PEP node.

Selection for growth in glucose medium of \emph{ppc-} mutants has
been studied via naturally evolved mutants,
resulting in the creation of two mutant strains \cite{fong2006latent} .
After 45 days of growth and selection, these mutants had growth rates
and glucose consumption rates very similar to the wild type strain, more
than double the \emph{ppc-} strain on day 0. While the underlying genetic
causes were not determined in this study, it was found that metabolic flux
through the glyoxylate shunt (aceA and aceB) increased and that there
was variable transcription of several genes, including decreased expression of
ptsH and ptsI, subunits of PEP-dependent phosphotransferase systems (PTS).
The glyoxylate shunt is an alternative route for the cell to make OAA.
Decreased expression of the PTS systems is likely to increase PEP
levels, as the cell will use sugar transport pathways that do not require
PEP. This is consitent with a study in which mutations in PTS were
found using a laboratory evolution setup selecting for increased growth
and succinate production \cite{Jantama08} . The mutations increased flux
through PEPCK in the non-native direction \cite{Jantama08} . In accordance with
Le Chatelier’s
principle, increasing the level of cellular PEP leads to higher
reverse PEPCK activity. Similarly, high PEP levels could also drive
flux through the glyoxylate shunt. Without extensive modeling of the
metabolic network of
the cell it is hard to predict \emph{a priori} which mutations
would be most beneficial to raising PEP levels.

As a more fundamental study of the paths that can circumvent
PPC, we designed a very simple selection, not focused
on succinate production, but rather looking at the many ways
that \emph{E. coli} can replenish OAA levels without PPC.
A $ppc^-$ strain cannot grow on glucose alone because there
is no anaplerotic supply of OAA to replenish the TCA
cycle. Unlike previous reports, we did not eliminate
pflB and therefore allowed for mutants that could supply
OAA through means other than PEPCK (for example through
TCA intermediates). We mutagenized the $ppc^-$ strain and
selected for growth on glucose. By performing 24 independent
mutagenesis and selection experiments we produced 24
separate mutant strains, identified mutations via
pooled sequencing, and identified genetic causes via
by pathway-phenoseq and gene-phenoseq.

\section{Results}
\label{paper:results}

\subsection*{Sequencing of Independent Mutants}
\label{paper:sequencing-of-independent-mutants}
Using growth on glucose medium as a selection, 24
mutants with the desired phenotype were produced. The
genomic DNA was pooled into 8 libraries each consisting
of exactly three strains. These libraries were tagged, combined,
and sequenced in a single lane of a high-throughput Illumina
Hi-Seq sequencer. The resulting fragments were filtered,
aligned to the reference \emph{E. coli} K-12 substr. MG1655 genome
sequence, and scanned for sequence variants.
Sequencing produced 145 million reads of 100
basepairs each for a total of 14.5 Gb of genomic sequence,
of which approximately 118 million reads successfully
demultiplexed (had an identifiable tag) and
aligned to the reference genome. From the pooled
libraries, we identified 2157 SNPS (1450 nonsynonymous,
707 synonymous) after filtering for quality and strand
bias (see methods). These SNPs showed a strong preference
for GC-sites in line with the mutagenesis spectrum of NTG
\cite{HarperNTG} . SNPs were detected in 1348 genes; 1012
genes had one or more nonsynonymous mutations.

\subsection*{Pathway-Phenoseq Analysis}
\label{paper:pathway-phenoseq-analysis}
We developed a method for scoring individual pathways, based on the
number of non-synonymous mutations occurring in genes in each
pathway (see Method for details).  As a comprehensive
set of \emph{E. coli} pathway annotations, we used the EcoCyc
Functionally Associated Groups database, totaling
536 groups \cite{EcoCyc11} , of which 336 were hit by
non-synonymous mutations in our sequencing dataset.
(For simplicity, we will refer to these EcoCyc Functionally
Associated Groups as ``pathways'').
We applied our scoring method
(which we will refer to throughout as ``pathway-phenoseq'')
to all 336 pathways, and ranked them by their p-values (Table 1).
After the Bonferroni multiple hypothesis correction, the
top five pathways (containing 12 genes total)
were statistically significant.

\begin{table}[!ht]
\caption{Top 10 gene groups ranked by pathway-phenoseq p-value (Bonferroni corrected for 536 tests)}

\begin{tabular}{|p{4cm}|p{4cm}|p{3.5cm}|}
\hline
\textbf{
Group
} & \textbf{
Genes
} & \textbf{
p-value (phenoseq)
}\\\hline

\textbf{PD04099}
 & 
\emph{aceK iclR}
 & 
$2.01 \times 10^{-39}$
\\\hline

\textbf{CPLX0-2101}
 & 
\emph{malE malF malG malK lamB}
 & 
$2.84 \times 10^{-9}$
\\\hline

\textbf{ABC-16-CPLX}
 & 
\emph{malF malE malG malK}
 & 
$7.17 \times 10^{-8}$
\\\hline

\textbf{PD00237}
 & 
\emph{malS malT}
 & 
$4.29 \times 10^{-4}$
\\\hline

\textbf{GLYCOGENSYNTH-PWY}
 & 
\emph{glgA glgB glgC}
 & 
$4.25 \times 10^{-3}$
\\\hline

\textbf{CPLX-155}
 & 
\emph{chbA chbB chbC ptsH ptsI}
 & 
$0.145$
\\\hline

\textbf{PWY0-321}
 & 
\emph{paaZ paaA paaB paaC paaD paaE paaF paaG paaH paaJ paaK}
 & 
$0.146$
\\\hline

\textbf{RNAP54-CPLX}
 & 
\emph{rpoA rpoB rpoC rpoN}
 & 
$0.53$
\\\hline

\textbf{APORNAP-CPLX}
 & 
\emph{rpoA rpoB rpoC}
 & 
$0.62$
\\\hline

\textbf{APORNAP-CPLX}
 & 
\emph{rpoA rpoB rpoC rpoD}
 & 
$0.71$
\\\hline
\end{tabular}

\end{table}

Overall, the top seven pathway-phenoseq scoring
pathways either have a clear connection to the PEP
metabolite level or offer
an alternative path of producing oxaloacetate (OAA),
and thus fit the $ppc^-$ growth phenotype (Figure 2).
PD04099 (\emph{aceK} and \emph{iclR}) regulates the glyoxylate shunt
\cite{Cozzone05} , \cite{Maloy82} , which is an alternative
way of producing OAA. By deregulating this pathway acetyl-CoA
can be used to regenerate OAA instead of being completely
consumed by the TCA cycle \cite{Kornberg57} . In fact iclR
has been deliberately mutated for biotechnology applications
to increase flux through the glyoxylate shunt and increase
yields of biomass or desired products \cite{waegeman2011effect} .
The next three association groups (CPLX0-2101, ABC-16-CPLX, and
PD00237) relate to the maltose transport pathway. If
deregulated the maltose transporter can transport glucose
using ATP as energy, instead of PEP like the PTS system
\cite{Boos88} . Therefore deregulation
of this group would increase PEP levels.
The fifth association group GLYCOGENSYNTH-PWY identified
involves mutations in the glycogen synthesis pathway.
Any mutations that decrease activity of this pathway
would increase glucose and ATP levels of the cell and
therefore decrease the PEP requirement of glucose transport.
These mutations are again correlated with increasing
cellular PEP levels. The chb (chitin PTS transport) pathway
CPLX-155 (association group six) is another sugar transport
pathway that requires PEP \cite{Keyhani00} . Any mutations
that downregulate or deactivate this pathway lead to
an increased pool of PEP in the cell, again consistent
with the selection performed. The genes \emph{ptsH} and \emph{ptsI} encode
general subunits of PTS systems that are required for
transport and phosphorylation of all sugars \cite{Saffen87} .
Therefore, PEP-dependent sugar transport in general is affected by
these mutations. In addition to hits in \emph{ptsHI}, mutations
were also found in \emph{crr}, the regulator responsible for
PEP-dependent transporters. We found mutations in sugar specific
subunits of several other PTS systems, however we would
need more data to boost the signal and determine which
are significant. The seventh association
group (PWY0-321) found in our selection encompasses genes
of the phenylacetate degradation pathway \cite{Teufel12} .
Like the first association group, this pathway answers
the selection not by increasing PEP levels, but instead
by supplying OAA through another TCA intermediate
(See Figure 2). The
final product of the phenylacetate degradation pathway
is succinyl-CoA, a TCA intermediate that can be metabolized
to OAA. All of the identified hits are relevant to OAA
levels, either by affecting PEP pools or by supplying direct
precursors of
OAA. Many of the hits identified here were not among
those that have been rationally explored by metabolic
engineers interested in the PEP/OAA node. This indicates
that there is great value in performing high throughput
evolution and sequencing experiments such as these to
identify new avenues of metabolic modifications relevant
to a pathway of interest. The final pathways in Table 1 consist of
rpoABC and rpoD and N genes which
affect gene expression globally. Mutations in these genes
can have widespread effects on gene expression. We do not
know the precise mechanisms of how these mutations may affect
the observed phenotype.
\begin{figure}[htbp]
\centering

\includegraphics[scale=0.4]{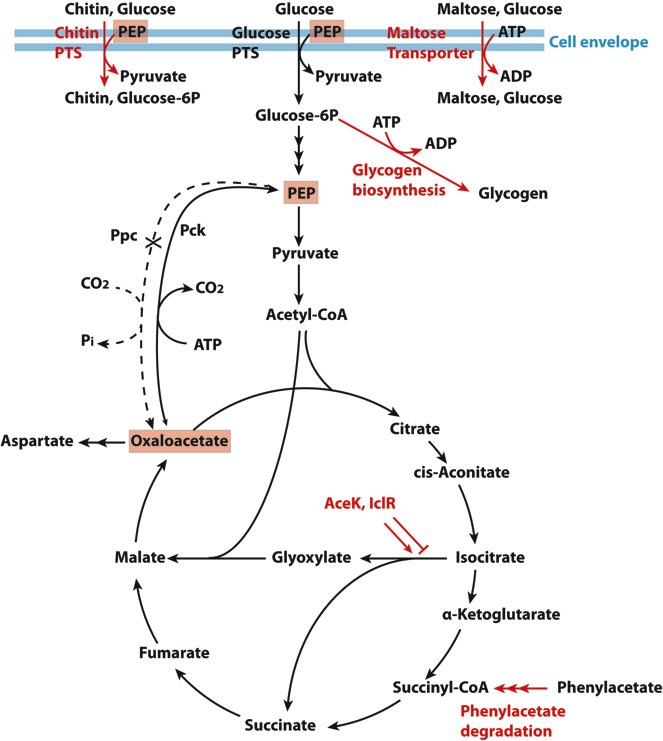}
\caption{\emph{Schematic of metabolic pathways affected by Ppc
knockout (dotted line). The :math:{}`text\{Ppc\}\textasciicircum{}-{}` strain requires
additional
oxaloacetate to grow. Growth is achieved through direct
synthesis of oxaloacetate by alternative pathways such as the
glyoxylate shunt or *pck}, or through an increase of PEP levels,
which drives flux through these pathways. The top seven
mutated pathways identified by pathway phenotype sequencing
are shown in red. It has been shown that Ppc knockouts cause increased
flux through
the glyoxylate shunt \cite{peng2006metabolic} , consistent with our
observed mutations in AceK and IclR. Mutations in PtsI have previously
been observed in response to a growth-based selection for increased
succinate production, in a scenario where Pck overexpression was also
observed \cite{zhang2009metabolic} . Similarly, deletion of ptsH,
which also deactivates the PTS system and increases the intracellular
PEP pool, has also been shown to increase succinate yields
\cite{zhang2009reengineering} .*}\end{figure}

\subsection*{Comparison with Gene-phenoseq}
\label{paper:comparison-with-gene-phenoseq}
By contrast, gene-phenoseq identified only
three of these genes (\emph{iclR} and \emph{aceK} in pathway
PD04099; \emph{malT} in pathway PD00237) as
statistically significant (Table 2).  Thus pathway-phenoseq
detected more than twice as many causal pathways
for this phenotype, and four times as many genes
as the gene-phenoseq scoring.

\begin{table}[!ht]
\caption{Top 20 hits ranked by Bonferroni corrected gene-phenoseq p-value computed on non-synonymous SNPs}

\begin{tabular}{|l|l|}
\hline
\textbf{
Gene
} & \textbf{
p-value
}\\\hline

\textbf{iclR}
 & 
$1.39 \times 10^{-25}$
\\\hline

\textbf{aceK}
 & 
$8.43 \times 10^{-14}$
\\\hline

\textbf{malT}
 & 
$4.81 \times 10^{-4}$
\\\hline

\textbf{malE}
 & 
$0.045$
\\\hline

\textbf{yjbH}
 & 
$0.088$
\\\hline

\textbf{rplL}
 & 
$0.18$
\\\hline

\textbf{ydfJ}
 & 
$0.18$
\\\hline

\textbf{pgi}
 & 
$0.21$
\\\hline

\textbf{yhcA}
 & 
$0.78$
\\\hline

\textbf{tyrS}
 & 
$0.82$
\\\hline

\textbf{yjaG}
 & 
$0.82$
\\\hline

\textbf{yeeN}
 & 
$0.82$
\\\hline

\textbf{tig}
 & 
$0.85$
\\\hline

\textbf{glgB}
 & 
$0.88$
\\\hline

\textbf{fdhF}
 & 
$0.89$
\\\hline

\textbf{gntT}
 & 
$1.04$
\\\hline

\textbf{dbpA}
 & 
$1.11$
\\\hline

\textbf{ydfl}
 & 
$1.16$
\\\hline

\textbf{lysC}
 & 
$1.18$
\\\hline

\textbf{xylE}
 & 
$1.22$
\\\hline
\end{tabular}

\end{table}

\subsection*{Bioinformatic Validation Tests}
\label{paper:bioinformatic-validation-tests}
As an additional test
of the entire set of top scoring pathways,
we computed a p-value for evidence of positive
selection (Ka/Ks \textgreater{} 1) within this set (Table 3).  Whereas
the phenoseq scoring is based on the \emph{total number}
of mutations in a region, the Ka/Ks is based
on the \emph{ratio} of non-synonymous vs. synonymous
mutations (note that the latter are \emph{not}
considered by the phenoseq scoring function).
The Ka/Ks ratio for the total dataset of 2157 SNPs
was 1.0026, consistent with neutral
selection, as expected from random mutagenesis.
We therefore computed a p-value for the null
hypothesis that mutations in the top pathways
are drawn from the same background distribution
as the total set of mutations (i.e. neutral)
using the Fisher Exact Test (see Methods for details).
The top 10 pathway-phenoseq pathways contained
a total of 103 non-synonymous mutations vs. only 21
synonymous mutations, yielding a p-value of
$3.38 \times 10^{-5}$.  This is strong evidence
of positive selection.  Even leaving out the genes
detected by gene-phenoseq (\emph{iclR}, \emph{aceK}, \emph{malT}),
the p-value is $5.12 \times 10^{-3}$.
Furthermore, this evidence of positive selection
extends throughout the top ten pathways.  For example,
if one leaves out pathways 6 through 10, the p-value
becomes weaker ($4.34 \times 10^{-4}$,
or again leaving out \emph{iclR}, \emph{aceK}, \emph{malT}, 0.056).
Indeed the p-value becomes \emph{stronger} (smaller p-value)
with each additional pathway added to the analysis, indicating that
each pathway shows evidence of positive selection.
Note that at the level of single-gene analysis,
only one gene (\emph{iclR} with 19 non-synonymous mutations
and 1 synonymous mutation) could be detected as showing
statistically significant evidence of positive
selection ($p=3.1 \times 10^{-3}$);
other genes simply did not have enough total
mutation counts to attain significance.

\begin{table}[!ht]
\caption{Positive Selection evidence for Top 10 gene groups}

\begin{tabular}{|l|l|l|}
\hline
\textbf{
Pathway
} & \textbf{
cumulative p-value
} & \textbf{
excluding \emph{iclR}, \emph{aceK}, \emph{malT}
}\\\hline

\textbf{PD04099}
 & 
0.0037
 & 
N/A
\\\hline

\textbf{CPLX0-2101}
 & 
0.0044
 & 
0.28
\\\hline

\textbf{ABC-16-CPLX}
 & 
0.0027
 & 
0.28
\\\hline

\textbf{PD00237}
 & 
0.0027
 & 
0.29
\\\hline

\textbf{GLYCOGENSYNTH-PWY}
 & 
0.0020
 & 
0.19
\\\hline

\textbf{CPLX-155}
 & 
0.00043
 & 
0.056
\\\hline

\textbf{PWY0-321}
 & 
0.000068
 & 
0.011
\\\hline

\textbf{RNAP54-CPLX}
 & 
0.000043
 & 
0.0063
\\\hline

\textbf{APORNAP-CPLX}
 & 
0.000043
 & 
0.0063
\\\hline

\textbf{APORNAP-CPLX}
 & 
0.000034
 & 
0.0051
\\\hline
\end{tabular}

\end{table}

It is interesting to ask what fraction of the genes
in these pathways show evidence of causing the phenotype.
It is evident (e.g. from the known experimentally validated
genes) that real causal genes are present far below the
0.05 significance threshold of gene-phenoseq scoring (also found
to be the case in a previous phenotype sequencing experiment
\cite{Harper2011} .)
To assess this, we took the top 50
gene-phenoseq genes, and asked what pathways were
strongly enriched (Table 4).
Given a top list of genes, one can assess whether
they cluster within specific subgroups of a standard
functional annotation using the hypergeometric
p-value test \cite{Fury2006} .  This analysis
identified statistically significant clustering
within three EcoCyc pathways.  Furthermore, six of the top ten
pathways matched the top 10 pathway-phenoseq pathways.
These data indicate that at least 9 of the genes in
these pathways contribute causally to the phenotype
(since they were individually detected among the top
50 gene-phenoseq hits). Only 28 pathways intersected the top 50
list.

\begin{table}[!ht]
\caption{Top 10 gene groups ranked by hypergeometric p-value (Bonferroni corrected for 28 tests)}

\begin{tabular}{|l|l|p{3cm}|p{3cm}||}
\hline
\textbf{
Group
} & \textbf{
Genes
} & \textbf{
Genes in top 20
} & \textbf{
p-value (hypergeometric)
}\\\hline

\textbf{ABC-16-CPLX}
 & 
\emph{malF malE malG malK}
 & 
$4$
 & 
$0$
\\\hline

\textbf{PD04099}
 & 
\emph{aceK iclR}
 & 
$2$
 & 
$0$
\\\hline

\textbf{CPLX0-2101}
 & 
\emph{malE malF malG malK lamB}
 & 
$4$
 & 
$6.875 \times 10^{-9}$
\\\hline

\textbf{CPLX-63}
 & 
\emph{torY torZ}
 & 
$1$
 & 
$0.0043$
\\\hline

\textbf{PD00237}
 & 
\emph{malS malT}
 & 
$1$
 & 
$0.0043$
\\\hline

\textbf{ABC-42-CPLX}
 & 
\emph{alsA alsB alsC}
 & 
$1$
 & 
$0.013$
\\\hline

\textbf{APORNAP-CPLX}
 & 
\emph{rpoA rpoB rpoC}
 & 
$1$
 & 
$0.013$
\\\hline

\textbf{GLYCOGENSYNTH-PWY}
 & 
\emph{glgA glgB glgC}
 & 
$1$
 & 
$0.013$
\\\hline

\textbf{SECE-G-Y-CPLX}
 & 
\emph{secE secG secY}
 & 
$1$
 & 
$0.013$
\\\hline

\textbf{CPLX0-221}
 & 
\emph{rpoA rpoB rpoC fecI}
 & 
$1$
 & 
$0.025$
\\\hline
\end{tabular}

\end{table}

\subsection*{Causal Mutations Analysis}
\label{paper:causal-mutations-analysis}
Finally, we sought to estimate the number of mutations
in each group that actually help cause the phenotype (``causal
mutations'').  In principle, one can estimate this from the
observed bias towards non-synonymous mutations (compared with
that expected under neutral selection as observed in the
total dataset).  Specifically, we assume that all causal
mutations must be non-synonymous, whereas non-causal
mutations are drawn from the background mixture of
synonymous + non-synonymous mutations (i.e. neutral
selection).  We can then estimate the fraction of
mutations in each pathway that are causal,
since the observed fraction of non-synonymous
mutations $f_o$ in a pathway will reflect the mix
$\theta$ of causal vs. non-causal mutations:
\[
f_o = \theta + (1-\theta)f_n
\]
where $f_n=1448/2157$ is the fraction of non-synonymous
mutations observed in the entire dataset (which almost
exactly matches that expected for neutral selection).  Then
\[
\theta = \frac{f_o-f_n}{1-f_n}
\]
We then estimated the number of causal mutations in
a pathway as $N_c = N\theta$, where \emph{N} is the total
number of mutations observed in the pathway (Table 5).
It is striking, for example, that the estimated number of
causal mutations in the top pathway (\emph{iclR} + \emph{aceK})
precisely equals the number of independent mutant strains
sequenced (24).  This suggests that each strain with
this phenotype was mutated once in this pathway.
The number of causal mutations estimated in the
remaining pathways ranged from 4 to 9, suggesting
that at least one additional mutation in these
other pathways was present in each strain. For each pool of three strains,
at least three nonsynonymous mutations were observed in the (\emph{iclR} + \emph{aceK})
pathway, so our data is consistent with the hypothesis that there
must be a mutation in this pathway to achieve the phenotype.

\begin{table}[!ht]
\caption{Estimated Causal Mutations in the Top 10 gene groups}

\begin{tabular}{|p{4cm}|p{3.5cm}|p{3.5cm}|p{3.5cm}|}
\hline
\textbf{
Group
} & \textbf{
Synonymous Mutations
} & \textbf{
Non-synonymous Mutations
} & \textbf{
Causal Mutations
}\\\hline

\textbf{PD04099}
 & 
5
 & 
34
 & 
24
\\\hline

\textbf{CPLX0-2101} / \textbf{ABC-16-CPLX}
 & 
6
 & 
18
 & 
6
\\\hline

\textbf{PD00237}
 & 
3
 & 
11
 & 
5
\\\hline

\textbf{GLYCOGENSYNTH-PWY}
 & 
3
 & 
10
 & 
4
\\\hline

\textbf{CPLX-155}
 & 
0
 & 
7
 & 
7
\\\hline

\textbf{PWY0-321}
 & 
1
 & 
11
 & 
9
\\\hline

\textbf{RNAP54-CPLX} / \textbf{APORNAP-CPLX} / \textbf{APORNAP-CPLX}
 & 
3
 & 
12
 & 
6
\\\hline
\end{tabular}

\end{table}

\section{Discussion}
\label{paper:discussion}
These data show first that pathway-phenoseq greatly improves
sensitivity and comprehensive discovery of the genetic
causes of a phenotype, over gene-phenoseq.  It detected
a statistically significant signal for more than
two times as many pathways, and an even greater proportion
of genes.  Second, our results indicate that independent
(non-competitive) mutant strains do indeed reveal a wide
variety of genetic causes of a phenotype, in this case:
regulators of the glyoxylate shunt; the maltose transport
pathway; the glycogen synthesis pathway; and the
phosphotransferase system.  Third, our analysis suggests
that the phenoseq approach is far more sensitive
for detecting such ``selection loci'' than standard
measures of selection such as Ka/Ks or dn/ds.
For example phenoseq detected a single pathway
with a p-value of $2 \times 10^{-39}$ (Bonferroni-
corrected), compared with a positive selection
p-value on the same pathway of 0.0037 (not even
Bonferroni-corrected).

We now consider some further implications and challenges
in this work.  First, it is evident that the number of
mutant strains sequenced both in this study (24) and
the previous isobutanol tolerance study (32) are
inadequate for definitively identifying all genes
that contribute to these phenotypes.  That is, our
results (and other experimental studies) have shown
clear evidence for a number of genes causing this
phenotype, that failed to attain statistical significance
in the gene-phenoseq scoring, e.g. PEPCK mutations
were not statistically significant in our study. In many cases
gene-phenoseq scoring ranked these genes highly, but
the sample size simply was not large enough to
yield a strong p-value.  This reflects a fact about
our two phenotypes which may apply generally to
many other phenotypes: they are complex, and involve
many genes, more than can be reliably detected by gene-phenoseq
with sequencing of 30 mutant strains.  It also illustrates
why pathway-phenoseq is needed: in our (admitedly limited)
experience, inadequate sensitivity is the key factor
limiting discovery.

Second, the same general approach should be applicable
to other mutagens and types of mutations.  As a minor
example, in the current study we did analyze promoter
mutations (separately from coding-region mutations),
but did not find any significant results (data not shown).
The same basic analysis should be applicable to
deletion mutations, transposon insert events and
any other mutational process for which one can
build an adequate neutral model.  In the worst case,
one could simply obtain data for a control set of
strains (i.e. mutated but not screened for the phenotype),
to provide an \emph{empirical} model of the mutational
bias of the set of genes, that would be used as
the null (non-target) model for scoring the
results observed after phenotype screening.

Third, it seems interesting to ask how many \emph{causal}
mutations are required to produce the phenotype.
We have presented a very simplistic way of estimating
the number of mutations in each pathway that are
actually causal.  This already seems to yield
intriguing suggestions, for example that the top
pathway (glyoxylate shunt regulation) is mutated
in essentially \emph{every} strain that has the phenotype,
and that this is typically accompanied by a ``second
mutation'' in another pathway.  It seems likely
that more sophisticated approaches to this question
will yield useful insights.  This is but one example
of exploring positive selection signals (concretely,
by taking synonymous mutations into account, which
phenoseq ignores).  Another possible application of
positive selection data is suggested by our Table 3:
whereas the total mutation dataset is unequivocally
neutral (Ka/Ks=1), the top-ranking pathways show
clear evidence of positive selection (Ka/Ks \textgreater{} 1).
Thus, it would be interesting to determine
(via a robust probabilistic analysis) how far
down the list of top-ranked phenoseq pathways
this positive selection signal goes (i.e.
where does it revert to neutrality).  In
principle, this could provide a measure of the
depth of phenotype selection in the dataset,
independent of the phenoseq p-value.

For the specific phenotype under study, we found that there
appears to be
a causal mutation in the regulatory pathway (iclR and aceK) of
the glyoxylate shunt (aceA and aceB) in every strain (Table 5).
Once activated, the shunt is then driven by a secondary mutation
in another pathway that alters the levels of PEP and/or OAA
in the cell. This is consistent with an earlier study
\cite{fong2006latent}  which produced two (naturally evolved) ppc-
mutants able to
grow in a glucose medium, both of which had increased flux
through the glyoxylate shunt as well as decreased expression
of ptsH or ptsI (our sixth ranked pathway), however the mutants
in that study were not sequenced, so it is not known which
genes were mutated. Moreover,
selection strength analysis (Table 3) shows that there
is evidence of positive
selection in pathways beyond the top phenotype sequencing
ranked pathway (iclR + AceK). It may be the case that
the top pathway is \emph{effectively larger} since it is potentially
more likely that a causal mutation occurs in a regulatory
gene than in an enzymatic gene (which may have a far smaller
active region). Regardless, a mutation in the regulatory pathway
is evidently not sufficient to induce the phenotype. Similarly,
it is possible that, in order to see PEPCK reversal, PEP levels
must be significantly elevated and pck must be deregulated. Our
experiment may not have contained a large enough sample size to
accommodate both of these requirements. In other studies the
elevated PEP levels are often artificially achieved through
pflB knockouts.

Finally, we must consider organisms where pathway
annotation is lacking (compared with the high level
of pathway annotation for \emph{E. coli}).  In principle,
any source of functional groupings of genes (for example
``Rosetta Stone'', phylogenetic profiles and related
non-homology approaches) could be used, in the absence
of human-curated pathway annotations.  Another interesting
possibility is to invert the problem: given a diverse set of
easily screenable phenotypes, one could systematically perform
phenotype sequencing on many such phenotypes, to obtain
observed groupings of genes that appear to ``function
together'' in the sense of causing the same phenotype(s).
Note that in contrast with a typical ``functional
correlation'' analysis (such as on expression levels),
even seeing a pair of genes as correlated by a single
data point (i.e. both causing one phenotype) would actually
be significant.  Thus far fewer phenotypes would have
to be studied to obtain significant results, than for
other functional correlation analyses such as expression levels.
Thus phenotype sequencing could itself be used as
a high-throughput method for finding functional groupings
of genes in less well studied microbial organisms.

\section{Methods}
\label{paper:methods}

\subsection*{Bacterial strains and growth conditions}
\label{paper:bacterial-strains-and-growth-conditions}
For strain construction and to prepare samples for NTG
mutagenesis strains were grown in standard Luria Bertani
medium \cite{Silhavy84} . Under selective conditions
strains were grown in a modified M9 medium (6 g
$\text{Na}_2\text{HPO}_4$,
3 g $\text{KH}_2\text{PO}_4$,
1 g $\text{NH}_4\text{Cl}$,
0.5 g NaCl,
1 mM $\text{MgSO}_4$,
1 mM $\text{CaCl}_2$,
10 mg vitamin B1 per liter of water) containing 1\% glucose.

Mutagenesis was performed on parent strain $ppc^-$. This
strain was generated by P1 transduction to delete \emph{ppc}
from \emph{E. coli} strain coli JCL16 (BW25113/F’ {[}traD36, proAB+,
lacIq ZDM15{]}) \cite{Atsumi2008b} , using strain JW3928
from the Keio collection as a P1 donor \cite{Baba06} .
This strain is unable to grow on glucose minimal medium.

\subsection*{NTG mutagenesis and selections}
\label{paper:ntg-mutagenesis-and-selections}
Random mutagenesis was performed with N’-nitro-N-nitrosoguanidine
(NTG) as previously described \cite{Miller72} . Briefly,
cultures of $ppc^-$ were grown to exponential phase in LB
medium, washed twice with 0.1M citrate buffer and then
concentrated two-fold by centrifugation and suspension
in 0.1 M citrate buffer (pH 5.5). Samples of 2mL were
exposed to N’-nitro-N-nitrosoguanidine (NTG) at a final
concentration of 50 mg/ml for 30 minutes at 37°C to reach
a percentage kill of approximately 50\%. The cells were
washed twice with 0.1 M phosphate buffer (pH 7.0) and
grown in LB for one hour. The cells were then challenged
by plating on glucose minimal medium and grown at 37°C
for 3 days. This procedure was performed on 24 separate
samples of $ppc^-$, each of which was plated separately
to ensure genetically distinct populations of mutants.
One colony from each separate NTG experiment was selected,
restreaked on selective medium plates to verify the phenotype
and then cultured in liquid medium to obtain genomic DNA.

\subsection*{DNA library preparation and sequencing}
\label{paper:dna-library-preparation-and-sequencing}
Bacterial genomic DNA was prepared from 24 mutant strains
using the DNEasy kit from
Qiagen using the optional RNAse treatments. The isolated
genomic DNA from the mutant strains was pooled in 8 pools,
each at a total concentration of 20ng/\(\mu\)L.
Equal amounts of DNA from 3 mutant strains were mixed in
each of the 8 pools.   The pooled samples were then fragmented
by sonication to an average size of 100–250 bp and confirmed
by gel electrophoresis. 8 tagged genomic sequencing libraries
(8 different indexes) were constructed using the TruSeq
DNA Sample Prep Oligo Kit following the low throughput
protocols provided by the manufacturer (Illumina). The
final concentration of each of the 8 indexed libraries
was measured by QuantiFluor assay and the 8 libraries
were mixed in equal proportion at a final concentration
of 10nM. 100bp single end read sequencing was carried
out on a single lane of an Illumina Genome Analyzer
HiSeq 2000 instrument
by the UCLA Broad Stem Cell Research Center High Throughput
Sequencing Facility.

\subsection*{Pathway-Phenoseq Analysis}
\label{paper:id1}
Short read data were aligned to the reference \emph{E. coli} genome
(Genbank accession NC\_000913) using Novoalign
(Novocraft, Selangor, Malaysia) in single-end mode.
Sequence variants were then called using samtools
\cite{LiSAM09}  mpileup and bcftools output to VCF format.
Only single nucleotide substitutions were found via
this analysis, consistent with NTG mutagenesis.
We then employed our phenoseq software package to
apply a succession of variant filters:
\begin{itemize}
\item {} 
we excluded variants with inadequate samtools quality scores.
Specifically, we required a QUAL value of greater than 90.

\item {} 
we excluded reported variants with strong evidence of
strand bias (i.e. the evidence for the variant came
primarily from reads in one direction but not the other).
Specifically, we excluded variants with a samtools AF1 p-value
of less than $10^{-2}$.  This eliminated a large
number of variant calls that appear to have been sequencing
errors.

\item {} 
we excluded variants with samtools allele frequency estimate
greater than 50\% in any given pool.  Concretely, each
independent mutant strain is expected to have different
mutations, so each mutation should be present in only
one out of three of the strains mixed together in one pool.

\item {} 
we excluded variants that were found in multiple tagged pools.
In all cases these were found in all 8 pools, indicating that
they were parental strain mutations (i.e. differences versus
the reference genome sequence).

\end{itemize}

We then used the EcoCyc functionally associated gene groups
to score pathways as follows:
\begin{itemize}
\item {} 
we only included non-synonymous mutations in the phenoseq analysis.
Specifically, we used the Pygr software package \cite{LeePygr09} 
to map the Genbank CDS annotations on the reference genome,
to map mutations to CDS (gene) intervals, and to
determine their effect on the amino acid translation.
Mutations that did not map to a CDS, or did not alter
the amino acid translation, were excluded.

\item {} 
CDS-mapped mutations were mapped to each EcoCyc group
using the EcoCyc database.

\item {} 
The expected mutational cross-section $\lambda$
for each EcoCyc
group was calculated based on its GC composition,
and the total density of all observed mutations
on GC sites vs. AT sites over the whole genome.

\item {} 
We computed a p-value for the null hypothesis that the
observed mutations $k_{obs}$ in an EcoCyc pathway
were obtained by random chance, under a Poisson model
\[
p(K\ge k_{obs}|\text{non-target}, \lambda)
= \sum_{K=k_{obs}}^\infty{\frac{e^{-\lambda}\lambda^K}{K!}}
\]
These calculations were performed with the \textbf{scipy.stats}
module \cite{Scipy2001}.

\item {} 
We applied a Bonferroni correction to this p-value
by multiplying by the total number of EcoCyc pathways
groups $N_e=536$.

\end{itemize}

We performed positive selection tests on these EcoCyc
pathway groups as follows:
\begin{itemize}
\item {} 
for a given set of one or more EcoCyc pathways, we obtained the counts
$N_a,N_s$ of non-synonymous vs. synonymous mutations
in that set of pathways.

\item {} 
We computed the p-value for obtaining this result under a
neutral (i.e. $K_a/K_s=1$), random model:
\[
p(m \ge N_a|n=N_a+N_s,N,M)
\]
where \emph{N} is the total number of all observed synonymous +
non-synonymous mutations in the whole genome, and \emph{M} is the
total number of observed non-synonymous mutations in the
whole genome.  Specifically, we computed this p-value
using the one-tailed (``greater'') Fisher Exact Test in R
\cite{Rmanual} .

\item {} 
Note that since only a single p-value test was performed
(on the top-ranked set of pathways), no Bonferroni correction
was applied.

\end{itemize}

Similarly, we computed p-values for pathway ``enrichment''
among the top 50 gene-phenoseq genes using the hypergeometric
test, again computed using the Fisher Exact Test in R or scipy \cite{Scipy2001} ,
with a Bonferroni correction corresponding to the number
of pathways that this test was applied to.

All of our code is available under an open source license
at {https://github.com/cjlee112/phenoseq}.

\section{Author Contributions}
\label{paper:author-contributions}
Conceived and designed the experiments: LG JCL MAH CJL.
Performed the experiments: LG JCL. Analyzed the data:
MAH CJL. Contributed reagents/materials/analysis tools:
MAH CJL JCL. Wrote the paper: MAH LG CJL.

\bibliographystyle{plain}
\bibliography{ref}

\end{document}